\documentclass[11pt]{article}
\usepackage{mathrsfs}
\usepackage{axodraw}

\newcommand{\Eq}[1]{Eq.\ (\ref{#1})}
\newcommand{\Eqs}[2]{Eqs.\ (\ref{#1}) and (\ref{#2})}
\newcommand{\Eqss}[3]{Eqs.\ (\ref{#1}),  (\ref{#2}) and (\ref{#3})}
\newcommand{\Ref}[1]{Ref.\ \cite{#1}}
\newcommand{\Fig}[1]{Fig.\ \ref{#1}}
\newcommand{\ms}{\mathscr}
\newcommand{\lslash}[1]{\rlap/#1}
\newcommand{\Tr}{\mathop{\rm Tr}}
\newcommand{\grav}{{\cal G}}

\newcommand{\nugmom}{\nu_1(p) \rightarrow \nu_2(p^\prime) + \grav(q)}

\title{\bf Angular momentum non-conserving decays in isotropic media}

\author{\bf Jos\'e F. Nieves\\
Laboratory of Theoretical Physics\\ 
Department of Physics, P.O. Box 23343\\
University of Puerto Rico, R\'{\i}o Piedras,
Puerto Rico 00931-3343
\and
\bf Palash B. Pal\\
Saha Institute of Nuclear Physics\\ 
1/AF Bidhan-Nagar, Calcutta 700064, India
}

\date{March 2009}

\evensidemargin=\oddsidemargin
\makeatletter
\@addtoreset{equation}{section}

\makeatother

\begin{document}

\maketitle

\begin{abstract}

  Various processes that are forbidden in the vacuum due to angular
  momentum conservation can occur in a medium that is isotropic and
  does not carry any angular momentum.  We illustrate this by
  considering explicitly two examples. The first one is the decay of a
  spin-0 particle into a photon and another spin-0 particle, using a
  model involving the Yukawa interactions of the scalar particles with
  a charged fermion field.  The second one involves the decay of a
  neutrino into another neutrino and a graviton, in the standard model
  of particle interactions augmented with the linearized gravitational
  couplings.

\end{abstract}

%
% $Id: amvdk_sec1.tex,v 1.1 2009/06/02 15:07:26 nieves Exp nieves $
%
% sec 1
%
\section{Introduction}
\label{s:introduction}

Some physical processes that are not allowed to occur in the vacuum
can occur in the presence of a background medium.  We can classify
such processes into two broad classes.  The first one refers to those
processes that are disallowed in the vacuum for kinematic reasons. One
classic example of this type is the Cerenkov radiation of a charged
particle, which can occur in a medium due to the fact that the photon
dispersion relation is modified by the background effects.  We are not
concerned with processes of this type here.

The second class consists of processes that do not occur because
the transition matrix element is zero in the vacuum. In general,
whenever that is the case, it can be attributed to some conservation laws,
which are in turn the consequences of the symmetries of the Lagrangian.
It is common to refer to such process as being \emph{forbidden}.
Thus, when the medium is not invariant under the full symmetry group
of the Lagrangian, the relevant transition matrix element can be
non-zero when the background effects are taken into account.

In a recent paper \cite{Nieves:2007jz} we considered a subset of
processes in this class, namely those which are forbidden
by helicity arguments, or angular momentum conservation.
We specifically considered the
radiative decay of a spin-0 particle into another spin-0 particle,
the decay of a spin-1 particle into two photons,
the gravitational decay of a spin-0 particle into another spin-0 particle
and the gravitational decay of a spin-1/2 particle into another spin-1/2.
By performing a form-factor analysis in each case, we reviewed the
arguments that show that the amplitude for the process in the vacuum vanishes,
and then demonstrated that the amplitude need not vanish if the process
occurs in a medium, even if the medium is homogeneous and isotropic and
therefore does not carry any net angular momentum.

The aim of the present work is to pursue this further to confirm that
this is indeed the case, by computing the amplitude for such
processes in viable models and verifying that it does
not vanish due to some unexpected reason. Here we consider the
radiative decay of a scalar particle and the gravitational decay of
a neutrino as illustrative examples of that kind of process.
In the scalar decay case, the model consists of two electrically
neutral scalar fields coupled to a charged fermion field
via Yukawa interactions, and the medium is assumed to consist
of a thermal background of the charged fermions. In the neutrino case,
the medium is a thermal background of electrons and,
in order to include the graviton interactions, 
the Standard Model couplings are supplemented
with the linearized gravitational couplings of the particles involved.
In either case, the thermal background is parametrized
by the Fermi-Dirac momentum distribution functions of the background
particles in the usual way.

The fact that these processes are forbidden in the vacuum
by angular momentum conservation arguments,
implies that, in the medium, their angular distribution and differential
decay rates have a distinctive form. This could lead to observable
consequences in specific physical contexts despite the fact that
there may exist other competing processes with comparable total rates.

In general, the presence of the medium modifies the
dispersion relations and wave function normalization of the particles
that participate in the process. Those corrections affect
the kinematics, but their relative importance depends on the particular
application and the physical context of the calculation.
Here we assume that the situation is such that those corrections
are negligible. However, the inclusion of those corrections
in the calculation of the rates is straightforward\cite{example},
and it should be kept in mind that they may be important
and need to be included in specific applications.

In Sec.\,\ref{s:raddk} we consider in detail the scalar radiative decay.
There we summarize the
the form-factor analysis presented in Ref.~\cite{Nieves:2007jz},
show that the on-shell amplitude depends only on one form factor,
and the expressions for the total and the differential decay rates
in terms of the form factor are given. We then carry out
the calculation of the amplitude in
a simple model involving a background of charged fermions, and the
one-loop formula for the on-shell form factor is obtained
in terms of integrals over the background fermion distribution functions.
The integrals are evaluated explicitly for some particular cases
of the distribution functions. The analogous calculations for the
flavor-changing gravitational decay of a neutrino are presented
in Sec.\,\ref{s:gravdk}. Sec.\,\ref{s:disc} contains some general
and concluding remarks.

%
% $Id: amvdk_sec2.tex,v 1.1 2009/06/02 15:07:26 nieves Exp nieves $
%
% sec2
%
\section{Radiative decay of a spinless particle}
\label{s:raddk}

\subsection{Kinematical considerations}
\label{subs:kin}

In this section, we consider the process 
\begin{eqnarray}
\phi(p) \rightarrow \phi'(p') + \gamma(q) \,,
\label{process}
\end{eqnarray}
where, $\phi$ and $\phi'$ denote the scalar (spin-0) particles, $\gamma$
denotes the photon, and $p$, $p'$ and $q$ denote the
corresponding momentum vectors. As mentioned in the Introduction,
we neglect the effects of the medium on the dispersion relations
and wave function normalizations in the calculation of the decay rate.
While it should be kept in mind that in general those corrections must be taken
into account in specific applications, they are not essential for our
purposes here. Moreover, they can be included in the calculations that follow
in a straightforward way if needed. Therefore for our purposes,
we assume the vacuum on-shell relations
\begin{eqnarray}
p^2 & = & m^2\,, \nonumber\\
p^{\prime\,2} & = & m^{\prime\,2}\,, \nonumber\\
q^2 & = & 0 \,.
\end{eqnarray}
The medium is assumed to consist of
a homogenous and isotropic thermal background of particles. Besides
the thermodynamic variables such as temperature and chemical potentials,
such a medium is characterized by its velocity four vector
$v^\mu$\cite{Weldon:1982bn,Weldon1983}.

The amplitude can be written in the form
\begin{eqnarray}
{\cal M} = \epsilon^{\ast\mu}(q) j_\mu \,,
%\label{}
\end{eqnarray}
where $\epsilon^\mu(q)$ is the photon polarization vector 
which satisfies
\begin{eqnarray}
q^\mu \epsilon_\mu(q) = 0\,,
\label{q.e}
\end{eqnarray}
and $j_\mu$ is the matrix element of the electromagnetic current,
which satisfies the transversality condition
\begin{eqnarray}
q^\mu j_\mu = 0 \,.
\label{q.j}
\end{eqnarray}

In general $j_\mu$ is a function of $p_\mu$ and $q_\mu$, which are
the only independent momenta in the problem.  When the
process takes place in a medium, $j_\mu$ can
also depend on $v_\mu$. In Ref.~\cite{Nieves:2007jz}, it was pointed out that
the most general form for the on-shell vertex function $j_\mu$
subject to the transversality condition of \Eq{q.j} is
\begin{eqnarray}
\label{jmu}
j_\mu = a \big[p\cdot q \, v^\mu - q \cdot v \, p^\mu\big] +
b \epsilon_{\mu\alpha\beta\gamma} p^\alpha q^\beta
v^\gamma\,. 
\end{eqnarray}
This implies that the decay amplitude can be written as
\begin{eqnarray}
{\cal M} = \Big( a F^*_{\mu\nu} + b \tilde F^*_{\mu\nu} \Big) v^\mu p^\nu\,, 
\label{Mwithab}
\end{eqnarray}
where 
\begin{eqnarray}
F_{\mu\nu} = \epsilon_\mu q_\nu - q_\mu \epsilon_\nu \,,
%\label{}
\end{eqnarray}
and $\tilde F_{\mu\nu} = \frac12 \varepsilon_{\mu\nu\alpha\beta}
F^{\alpha\beta}$ is the dual.  The form factors $a$ and $b$ appearing
in the amplitude are Lorentz invariant functions of $p$, $q$ and $v$. 

The term associated with the form factor $b$ is parity violating due
to the presence of the Levi-Civita tensor.  In a system in
which the interactions are parity conserving and the background
medium is parity symmetric, i.e., is not spin-polarized, the form
factor $b$ is zero..  In the calculation that we carry out
subsequently in this paper, we assume that this is the case,
and therefore we set $b=0$ and write
\begin{eqnarray}
\label{M}
{\cal M} = a F^*_{\mu\nu} v^\mu p^\nu \,.
\end{eqnarray}
Without loss of generality, we will take the form factor $a$ to be real.

Choosing the convention in which the $z$-axis
points in the direction of the motion of the decaying particle,
the components of the vectors $v^\mu$ and $p^\mu$ in the rest frame
of the medium are
\begin{eqnarray}
v^\mu = (1, \vec 0)\,, \qquad 
p^\mu = (E, P \hat z)\,,
\end{eqnarray}
and we will denote by
\begin{equation}
V = P/E
\end{equation}
the velocity of the decaying particle. The differential decay rate is then
\begin{eqnarray}
\label{dGdtheta}
\frac{d\Gamma}{d(\cos\theta)} = \frac{\omega_0}{16\pi m^2}
\frac{(1 - V^2)^{3/2}}{(1 - V\cos\theta)^2}\overline{|{\cal M}|^2}\,,
\end{eqnarray}
where $\theta$ is the angle between $\vec q$ and the $z$-axis,
\begin{eqnarray}
\label{omega0}
\omega_0 = \frac{m^2 - m^{\prime\,2}}{2m} \,,
\end{eqnarray}
and 
\begin{equation}
\overline{|{\cal M}|^2} = \sum_{\rm pol} |{\cal M}|^2 \,,
\end{equation}
with the sum being over the two polarization states of the photon.
It may be convenient to express the differential rate in
terms of the photon energy, which is given by
\begin{eqnarray}
\label{gravitonomega}
\omega = \frac{m^2 - m^{\prime\,2}}{2E(1 - V\cos\theta)} \,.
\end{eqnarray}
Thus,
\begin{eqnarray}
\frac{d\Gamma}{d\omega} = \frac{1}{16\pi m^2}\frac{(1 - V^2)}{V}
\overline{|{\cal M}|^2}\,,
\end{eqnarray}
and the total rate is 
\begin{eqnarray}
\label{totalrateformula}
\Gamma = \frac{1}{16\pi m^2}
\frac{(1 - V^2)}{V}\int_{\omega_0 r}^{\omega_0/r}
d\omega \; \overline{|{\cal M}|^2}\,,
\end{eqnarray}
where
\begin{eqnarray}
r = \sqrt{\frac{1 - V}{1 + V}}\,,
\end{eqnarray}
and $\omega_0$ has been defined in \Eq{omega0}.

For the present case, in which the amplitude is given by \Eq{M},
\begin{eqnarray}
\label{Msq}
\overline{|{\cal M}|^2} &=& a^2 P^2 \omega^2 \sum_{\rm pol} |\vec
\epsilon\cdot \hat z|^2 \nonumber\\ 
&=& a^2 P^2 \omega^2 \sin^2 \theta  \,,
\end{eqnarray}
where we should remember that $\omega$ and $\theta$ are related
by \Eq{gravitonomega}. Thus, using \Eqss{dGdtheta}{omega0}{gravitonomega}
\begin{eqnarray}
{d\Gamma \over d(\cos\theta)} 
&=& {a^2 \over 16 \pi} \left( m^2 - m'^2 \over 2m \right)^3
V^2(1 - V^2)^{3/2}
\frac{\sin^2\theta}{(1 - V\cos\theta)^4}\,.
\label{diffGamma}
\end{eqnarray}

The integration that remains to obtain the total rate,
either from \Eq{diffGamma} or \Eq{totalrateformula},
cannot be performed until we have
the explicit formulas for the form factor, since in general it depends
on $\theta$ or, equivalently, $\omega$.
Here we observe that, since the functional dependence of the form
factor on $\theta$ will be different depending on conditions of the
fermion background, so will be the angular distribution.

%%%%%%%%%%
\subsection{The model and the diagrams}
\begin{figure}
\begin{center}
%
% $Id: amvdk_fig1.tex,v 1.1 2009/06/02 15:07:26 nieves Exp nieves $
%
\begin{picture}(120,120)(-60,-70)
\SetWidth{1.2}
\Text(0,-70)[]{\Large\bf (a)}
\ArrowArc(0,0)(30,-30,90)
\Text(27,16)[bl]{$l+q$}
\ArrowArc(0,0)(30,90,210)
\Text(-27,16)[br]{$l$}
\ArrowArc(0,0)(30,210,330)
\Text(0,-30)[t]{$l+p$}
\Photon(0,30)(0,55)24
\DashArrowLine(-60,-40)(-24,-16)3
\DashArrowLine(24,-16)(60,-40)3
\end{picture}
\quad \quad 
\begin{picture}(120,120)(-60,-70)
\SetWidth{1.2}
\Text(0,-70)[]{\Large\bf (b)}
\ArrowArc(0,0)(30,150,270)
\Text(27,-16)[tl]{$l-q$}
\ArrowArc(0,0)(30,-90,30)
\Text(-27,-16)[tr]{$l$}
\ArrowArc(0,0)(30,30,150)
\Text(0,30)[b]{$l-p$}
\Photon(0,-30)(0,-55)24
\DashArrowLine(-60,40)(-24,16)3
\DashArrowLine(24,16)(60,40)3
\end{picture}
\end{center}
\caption[]{One-loop diagrams for the process of \Eq{process}
in a thermal background of fermions.
Each line is labeled with its corresponding momentum variable.
\label{f:diags}}
\end{figure}
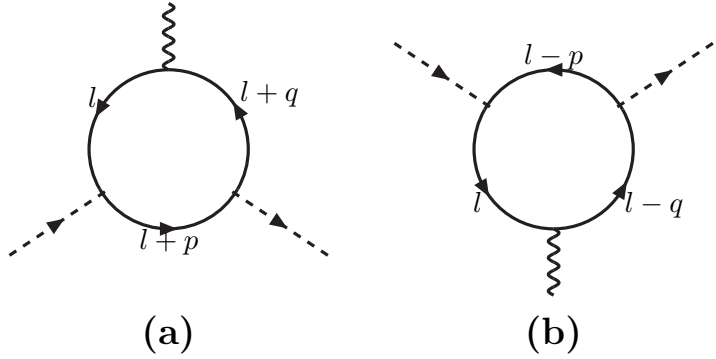

%%%%%%%%%%
In order to perform calculations to evaluate the form factors
we consider a model containing two neutral scalar fields $\phi$ and $\phi'$,
and a charged fermion $f$, with the interaction Lagrangian
\begin{eqnarray}
\ms L_Y = - \lambda \bar ff \phi - \lambda' \bar ff \phi' \,,
\label{LYuk}
\end{eqnarray}
in addition to the standard electromagnetic coupling of $f$.
The medium is assumed to be a thermal background of the fermions $f$.
The lowest order diagrams are shown in Fig.~\ref{f:diags},
where the internal line in the loops represent the fermions in
the thermal background. We write the fermion thermal propagator in the form
\begin{eqnarray}
\label{Se}
iS(l) = i(\rlap/l + m_f) T(l) \,,
\end{eqnarray}
where
\begin{eqnarray}
T(l) = \Delta(l) - 2\pi i \delta(l^2-m_f^2) \eta(l) \,,
%\label{}
\end{eqnarray}
with
\begin{eqnarray}
\Delta(l) = \frac{1}{l^2 - m_f^2 + i\epsilon} \,,
\end{eqnarray}
and
\begin{eqnarray}
\label{etae}
\eta(l) = {\Theta(l\cdot v) \over e^{\beta(l\cdot v - \mu_f)} + 1} + 
{\Theta(-l\cdot v) \over e^{-\beta(l\cdot v - \mu_f)} + 1} \,,
\end{eqnarray}
$\beta$ and $\mu_f$ being the inverse temperature and chemical potential
of the fermion gas, respectively.

The contributions to $j^\mu$ from the two diagrams are given by
\begin{eqnarray}
j_\mu^{(a)} &=& i4e_f \lambda \lambda' \int {d^4l \over (2\pi)^4} 
L_\mu(l+q,l+p,l) T(l+q) T(l+p) T(l) ,
\label{jmua}\\ 
j_\mu^{(b)} &=& i4e_f \lambda \lambda' \int {d^4l \over (2\pi)^4}
L_\mu(l, l-p, l-q) T(l) T(l-p) T(l-q) \,,
\label{jmub}
\end{eqnarray}
where $e_f$ is the electric charge of the fermion $f$, and
\begin{eqnarray}
L_\mu(p_1,p_2,p_3) & \equiv & \frac{1}{4}\Tr\left[\gamma_\mu
  (\rlap/p_1 +m_f )   (\rlap/p_2 + m_f) (\rlap/p_3 + m_f)\right] \,,\nonumber\\
& = & p_{1\mu} p_2\cdot p_3 - p_{2\mu} p_3\cdot p_1 + p_{3\mu}
  p_1\cdot p_2 + (p_{1\mu} + p_{2\mu} + p_{3\mu})m_f^2 \,.
\label{Ldef}
\end{eqnarray}
This expression confirms the statement made in Sec.\ \ref{subs:kin},
that the form factor $b$ that appears in \Eq{jmu} vanishes in this model.
As stated there, this is a consequence of parity invariance of the
interactions of \Eq{LYuk} and the polarization independence of the
fermion distribution functions in the medium.

Using the relationships
\begin{eqnarray}
L_\mu(p_1,p_2,p_3) & = & L_\mu(p_3,p_2,p_1) \,,\nonumber\\
L_\mu(-p_1,-p_2,-p_3) & = & -L_\mu(p_1,p_1,p_3) \,,
%\label{}
\end{eqnarray}
\Eqs{jmua}{jmub} can be combined in the form
\begin{eqnarray}
\label{j}
j_\mu &=& i4e_f \lambda \lambda' \int {d^4l \over (2\pi)^4} 
L_\mu(l+q,l+p,l) \nonumber\\* 
&& \times \left[T(l+q) T(l+p) T(l) - T(-l-q) T(-l-p) T(-l)\right] \,. 
\end{eqnarray}
When \Eq{Se} is substituted into \Eq{j} various terms are produced.
The purely vacuum terms, which do not contain any factor involving
$\eta(l)$, give zero since $\Delta(-l) = \Delta(l)$.  The terms
containing three factors of $\eta$ also yield zero due to the on-shell
conditions implied by the delta function in \Eq{etae}.  The terms
containing two factors of $\eta$ will give an absorptive contribution
to the amplitude. Here we will assume that the kinematic regime is
such that those absorptive terms are zero, i.e., that the initial
$\phi$ state is below the $f\bar f$ pair production threshold. Therefore the
only the terms that survive are those that contain one factor of
$\eta$, and by making appropriate shifts of the integration variables
in some of those terms, we obtain
\begin{eqnarray}
j_\mu  &=&  4e_f \lambda \lambda' \int {d^4l \over (2\pi)^3} \delta(l^2-m_f^2)
[\eta(l) - \eta(-l)]  \Big[L_\mu(l+q,l+p,l)\Delta(l+p) \Delta(l+q)
\nonumber\\*
&& + L_\mu(l-p+q,l,l-p)\Delta(l-p)\Delta(l-p+q)\nonumber\\*
&& + L_\mu(l,l+p-q,l-q) \Delta(l-q)\Delta(l+p-q) \Big] \,.
\label{jfinal}
\end{eqnarray}
%

%%%%%%%%%%
\subsection{Transversality condition}
%%%%%%%%%%
The transversality condition $q_\mu j^\mu = 0$ is easily verified
explicitly with the help of the identity
\begin{eqnarray}
(p_1 - p_3)^\mu L_\mu(p_1,p_2,p_3) = (p_2\cdot p_3 + m_f^2)\Delta^{-1}(p_1) -
(p_2\cdot p_1 + m_f^2)\Delta^{-1}(p_3) \,,
\end{eqnarray}
which follows simply from \Eq{Ldef}. Using this identity to rewrite
the various factors of $q^\mu L_\mu$ that appear when we contract
\Eq{jfinal} with $q^\mu$, we obtain
\begin{eqnarray}
\label{qdotj1loop}
q^\mu j_\mu & = & 4e \lambda \lambda'
\int {d^4l \over (2\pi)^3} \delta(l^2-m_f^2)
[\eta(l) - \eta(-l)]\left\{\Delta(l+p)[l\cdot(l + p) + m_f^2]\right.
\nonumber\\
&&\mbox{} + \Delta(l - p)[l\cdot(l - p) + m_f^2] -
\Delta(l - p + q)[l\cdot(l - p + q) + m_f^2]\nonumber\\ 
&&\mbox{} - \left. \Delta(l + p - q)[l\cdot(l + p - q) + m_f^2]\right\} \,,
\end{eqnarray}
where we have also used the relation
\begin{eqnarray}
\delta(l^2 - m_f^2)\Delta^{-1}(l^2 - m_f^2) = 0\,.
\end{eqnarray}
Since the factor $\eta(l) - \eta(-l)$ in the integrand of \Eq{qdotj1loop}
is odd under $l \rightarrow -l$, while the other factor, within
the curly brackets, is even, the integrand is odd and therefore \Eq{q.j}
is verified.

%%%%%%%%%%
\subsection{Evaluation of the form factor}
%%%%%%%%%%
Using the trace formula of \Eq{Ldef}, we can write down the individual
terms in $j_\mu$ from \Eq{jfinal}.  For this, it is convenient to
define the integrals
\begin{eqnarray}
I_\alpha (p_1,p_2) &\equiv& \int_l  
{l_\alpha \over (p_1^2 + 2l\cdot p_1)(p_2^2 + 2l\cdot p_2)} \,,
\label{Ialpha}\\
I (p_1,p_2) &\equiv& \int_l 
{1 \over (p_1^2 + 2l\cdot p_1)(p_2^2 + 2l\cdot p_2)} \,,
\label{I} 
% \\
% K_\alpha (p_1) &\equiv& \int {d^4l \over (2\pi)^3} \delta(l^2-m^2)
% [\eta(l) - \eta(-l)] {l_\alpha \over (p_1^2 + 2l\cdot p_1)} 
% \label{Kalpha} \,,\\
% K (p_1) &\equiv& \int {d^4l \over (2\pi)^3} \delta(l^2-m^2)
% [\eta(l) - \eta(-l)] {1 \over (p_1^2 + 2l\cdot p_1)} \,.
% \label{K}
\end{eqnarray}
where
\begin{eqnarray}
\label{Il}
\int_l \equiv \int {d^4l \over (2\pi)^3} \delta(l^2-m_f^2)
[\eta(l) - \eta(-l)] \,.
\end{eqnarray}
These integrals satisfy the relations
\begin{eqnarray}
I_\alpha(-p_1,-p_2) &=& I_\alpha(p_1,p_2) \,, \\
I(-p_1,-p_2) &=& - I(p_1,p_2)\,.
% \,, \\
% K_\alpha(-p_1) &=& K_\alpha(p_1) \,, \\
%K(-p_1) &=& - K(p_1) 
%\label{}
\end{eqnarray}
In addition, since $q^2=0$, both integrals are odd in $q$ if either $p_1 = q$
or $p_2 = q$. Using these properties, we
can write the vertex function in the form
\begin{eqnarray}
j_\mu &=& 4e_f\lambda\lambda' \bigg(
(4m_f^2 - p \cdot p') \bigg[ 
  I_\mu(p,p') + I_\mu(p,q) - I_\mu(p',q) \bigg] \nonumber\\*
&& + p_\mu \bigg[ (p+p')_\alpha I^\alpha(p,p') - q_\alpha
I^\alpha (p,q) - q_\alpha
I^\alpha (p',q) + 4m_f^2 I(p,p') \bigg] \bigg) \,,
\label{jI}
\end{eqnarray}
where we have omitted terms proportional to $q_\mu$
that do not contribute to the amplitude due to \Eq{q.e}.
In order to determine the form factor $a$ from this expression,
we only need to look for the terms proportional to $v_\mu$,
according to \Eq{jmu}. Such terms can arise only
from the integrals $I_\mu$, which can be combined in the form
\begin{eqnarray}
I_\mu(p,p') + I_\mu(p,q) - I_\mu(p',q) = - (m^2-m'^2) \tilde
I_\mu(p,p') \,, 
\end{eqnarray}
where
\begin{eqnarray}
\label{Itildedef}
\tilde I_\mu (p,p') = \int_l {l_\mu
  \over 2l\cdot q (m^2 + 2 l\cdot p) (m'^2 + 2 l\cdot p')} \,.
\end{eqnarray}
The result of the integration can be expressed in the form
\begin{eqnarray}
\tilde I_\mu (p,p') = A v_\mu + B p_\mu + B' p'_\mu \,,
\label{Itilde}
\end{eqnarray}
where $A$, $B$ and $B'$ are Lorentz invariants, and the form factor $a$
is then determined as
\begin{eqnarray}
a = 4e\lambda\lambda' (m^2 + m'^2 - 8m_f^2) A \,.
%\label{}
\end{eqnarray}

The integral defined in \Eq{Itildedef} cannot be calculated analytically
in the general case, as is common for this type of integral
that involve the thermal distribution functions. For definiteness
here we consider the case of a non-relativistic fermion gas
for which a simple result can be obtained. Thus, assuming that
\begin{eqnarray}
|l_0| \approx m_f, \qquad |\vec l| \ll m_f \,,
%\label{}
\end{eqnarray}
then as a first approximation we can neglect $\vec l$ altogether and
replace $l_0$ by $m_f$ in \Eq{Itilde}. As a result, $\tilde I_\mu$
is proportional to $v_\mu$, so that the coefficients $B$ and $B'$
are zero, while
\begin{eqnarray}
A = {1 \over 2\omega (m^2 + 2 m_f E) (m'^2 + 2m_f E')} \int {d^4l \over
  (2\pi)^3} \delta(l^2-m_f^2) [\eta(l) - \eta(-l)] \,,
%\label{}
\end{eqnarray}
where
\begin{equation}
E^\prime = E - \omega \,.
\end{equation}
Therefore,
\begin{eqnarray}
a &=&  e_f\lambda\lambda' (n_f - n_{\bar f})  
{m^2 + m'^2 - 8m_f^2 \over 2m_f \omega(m^2 + 2m_f E)(m'^2 + 2m_f E')} \,,
\label{aNR}
\end{eqnarray}
where $n_f$ and $n_{\bar f}$ are the number densities of the fermions $f$
and their antiparticles, respectively. 

If the background is charge-symmetric, i.e.,
contains an equal number of particles and antiparticles,
the form factor, and thereby the amplitude, is zero as \Eq{aNR}
clearly shows. This result actually follows more generally from the presence
of the factor $\eta(l)-\eta(-l)$ in \Eq{Il}, and it holds
whether the fermion gas is non-relativistic or not.
On the other hand, \Eq{aNR} confirms explicitly that the amplitude is
non-zero in general, giving a non-zero rate for the radiative decay process.

Finally, the explicit expression for the differential and total decay
rates in this limit (non-relativistic gas) are given by substituting 
\Eq{aNR} into \Eqs{totalrateformula}{diffGamma}, and remembering
the relation given in \Eq{gravitonomega}. The final integration that
remains to obtain the total rate using either
\Eq{diffGamma} or \Eq{totalrateformula},
is straightforward but cumbersome. We do not pursue this further
here since the details are similar to next case that we consider,
which we treat in full.
 
%
% $Id: amvdk_sec3.tex,v 1.1 2009/06/02 15:07:26 nieves Exp nieves $
%
% sec 3
%
\section{Gravitational decay of a neutrino}
\label{s:gravdk}
\subsection{Kinematical considerations}

We now consider the process
\begin{eqnarray}
\label{s2:process}
\nugmom
\end{eqnarray}
where $\grav$ denotes the graviton.  The amplitude can be written as
\begin{eqnarray}
\label{s2:physicalamplitude}
i{\cal M} = -i\kappa\epsilon^{\lambda\rho\ast}(q)
\bar u_2(p^\prime)\Gamma_{\lambda\rho}u_1(p)\,,
\end{eqnarray}
where $\epsilon^{\lambda\rho}$ is the polarization tensor of the
graviton, $\Gamma_{\lambda\rho}$ is the vertex function, and $\kappa$
is related to the Newton gravitational constant $G$ by
\begin{eqnarray}
\kappa = \sqrt{8\pi G}\,.
\end{eqnarray}
As we have already mentioned, in the calculation of the rate we
neglect the effects of the medium in the dispersion relations
and wave function renormalization of the external particles.
Therefore, the spinors satisfy the vacuum Dirac equation while the
on-shell conditions for the graviton polarization tensor are
\begin{eqnarray}
\label{gravonshell}
\epsilon_{\lambda\rho}(q) & = & \epsilon_{\rho\lambda}(q)\,,\nonumber\\
\eta^{\lambda\rho} \epsilon_{\lambda\rho}(q) & = & 0 \,,\nonumber\\
q^\lambda\epsilon_{\lambda\rho}(q) & = & 0\,.
\end{eqnarray}

The vertex function is constrained by Lorentz invariance and must be
transverse to the graviton momentum.  Subject to these conditions, the
general form for the vertex function has been obtained in 
\Ref{Nieves:2007jz}. It was noted there that the vertex function
can be decomposed into a tensor and a pseudotensor,
\begin{eqnarray}
\Gamma_{\lambda\rho} = \Gamma_{\lambda\rho}^{(T)} +
\Gamma_{\lambda\rho}^{(P)} \gamma_5 \,,
%\label{}
\end{eqnarray}
and that for each $\Gamma_{\lambda\rho}^{(I)}$ with $I=T,P$, the terms
that contribute to the vertex function with an on-shell graviton are
of the form
\begin{eqnarray}
\Gamma_{\lambda\rho}^{(I)} &=& \left[
-\frac{q\cdot v}{p\cdot q}p_\lambda p_\rho -\frac{p\cdot q}{q\cdot
  v}v_\lambda v_\rho + \{pv\}_{\lambda\rho} 
\right](a^{(I)}_5 + b^{(I)}_5\lslash{v}) \nonumber\\*
&& \null + \left[
\frac{q\cdot v}{(p\cdot q)^2}\lslash{q}p_\lambda p_\rho -
\frac{\lslash{q}}{q\cdot v}v_\lambda v_\rho -
\frac{q\cdot v}{p\cdot q}\{p\gamma\}_{\lambda\rho} +
\{v\gamma\}_{\lambda\rho} 
\right](a^{(I)}_9 + b^{(I)}_9\lslash{v})\,,
%\label{}
\end{eqnarray}
where the $a$'s and the $b$'s are form factors, and 
\begin{eqnarray}
\{AB\}_{\lambda\rho} = A_\lambda B_\rho + A_\rho B_\lambda \,.
%\label{}
\end{eqnarray}

For the present calculation, the on-shell amplitude takes a simpler form.
First, the transversality of the vertex function implies that
we can make the replacement
\begin{eqnarray}
\epsilon_{\lambda\rho} \rightarrow \epsilon_{\lambda\rho} +
X_\lambda q_\rho + q_\lambda X_\rho
\end{eqnarray}
with any $X_\lambda$,
which allows us to choose the graviton polarization tensor such that
\begin{eqnarray}
v^\lambda \epsilon_{\lambda\rho} = 0 \,.
%\label{}
\end{eqnarray}
Second, since the neutrinos are much
lighter compared to the charged leptons and the weak gauge bosons, we
can neglect the masses in calculating the leading term of the
amplitude.  This also enables us to use the massless limit formulas
\begin{eqnarray}
\label{nuonshell}
\bar u_2(p^\prime) \lslash{q} u_1(p) & = & 0 \,,\nonumber\\
\bar u_2(p^\prime) \lslash{q}\gamma_5 u_1(p) & = & 0 \,.
\end{eqnarray}
Finally, chirality arguments dictate that the terms with or without
the $\gamma_5$ in the amplitude are not independent. In fact, in the
massless limit, the chirality invariance of the standard model
interactions imply that the terms $a^{(i)}_5$ and $b^{(i)}_9$ cannot
be present, while the remaining ones actually appear combined in the form
\begin{eqnarray}
\label{chiralamplitudeonshell}
i{\cal M} = -i\kappa \varepsilon^{\ast\mu\nu}
\bar u_2(p^\prime)\left[c p_\mu p_\nu \lslash{v} +
d\left(p_\mu\gamma_\nu + p_\nu\gamma_\mu\right)\right]L u_1(p) \,,
\end{eqnarray}
where $c$ and $d$ independent coefficients that must be computed.
Eq.\ (\ref{chiralamplitudeonshell}) is the most general form of the
on-shell amplitude in the chiral limit.

The formulas already given in \Eqs{dGdtheta}{totalrateformula} for
the angular distribution and the decay rate apply in this case as well.
In order to evaluate $\overline{|{\cal M}|^2}$ we use the
relation 
\begin{eqnarray}
u_1\bar u_1 = L\lslash{p} \,,
\end{eqnarray}
which is satisfied by the neutrino spinor in the massless
approximation that we have used, and the analogous relation for $u_2$.
The graviton polarization tensor can be expressed in terms of the
spin-1 polarization vector of definite helicity $\epsilon^\mu_s$ (with
$s = \pm$), which are such that
\begin{eqnarray}
\epsilon^\mu_s = (0, \vec \epsilon_s) \,,
\end{eqnarray}
with
\begin{eqnarray}
\vec \epsilon_s\cdot\vec q = 0 \,.
\end{eqnarray}
A particularly convenient representation follows by defining the unit vectors
\begin{eqnarray}
\label{spin1rep}
\hat q & = & (\sin\theta\cos\phi, \sin\theta\sin\phi, \cos\theta)\,,
\nonumber\\
\vec e_1 & = & (\sin\phi, -\cos\phi, 0)\,,\nonumber\\
\vec e_2 & = & (\cos\theta\cos\phi, \cos\theta\sin\phi, -\sin\theta)\,.
\end{eqnarray}
The spin-1 polarization vectors of definite helicity are given by
\begin{eqnarray}
\vec \epsilon_s = \frac{1}{\sqrt{2}}(\vec e_1 + is\vec e_2) \,,
\end{eqnarray}
and the graviton polarization tensor for a definite helicity is
\begin{eqnarray}
\epsilon^{\mu\nu}_s = \epsilon^\mu_s \epsilon^\nu_s \,.
\end{eqnarray}

In the most general case, the coefficients $c$ and $d$ introduced
in \Eq{chiralamplitudeonshell} can have different phases, which
is a signal of $CP$-nonconserving effects. However, as is well known,
such effects do not show up at one-loop when their only source is
the mixing matrix. Thus, for our present purposes, we can assume that $c$
and $d$ are relatively real. The generalization to the the case in which
$\mbox{Im}\,c^\ast d \not= 0$ is straightforward. In this way we then obtain
\begin{eqnarray}
\label{Msquared}
\overline{|{\cal M}|^2} = 2\kappa^2 P^4 \sin^4\theta\left\{
|c|^2 E (E - \omega) + 4|d|^2 + 2c^\ast d(2E - \omega)\right\}
\end{eqnarray}
where we have used the relations
\begin{eqnarray}
\label{polsums}
\sum_s |\vec\epsilon_s\cdot\hat z|^2 &=& \sin^2\theta \nonumber\\
\sum_s |\vec\epsilon_s\cdot\hat z|^4 &=& 
\frac{1}{2}\sin^4\theta \,.
\end{eqnarray}

\subsection{The diagrams}
\label{s:vertexdiag}
\begin{figure}
\begin{center}
%
% $Id: amvdk_fig2.tex,v 1.1 2009/06/02 15:07:26 nieves Exp nieves $
%
% graviton coupled to electron
%
\begin{picture}(160,130)(-80,-65)
\SetWidth{1.2}
\Text(35,-35)[ct]{(a)}
\ArrowLine(80,0)(40,0)
\Text(60,-10)[c]{$\nu_e(p)$}
\ArrowLine(40,0)(0,0)
\Text(20,-10)[c]{$e(l)$}
\ArrowLine(0,0)(-40,0)
\Text(-5,-10)[r]{$e(l - q)$}
\ArrowLine(-40,0)(-80,0)
\Text(-60,-10)[c]{$\nu_e(p^\prime)$}
\Photon(0,0)(0,-45){2}{4}
\Photon(0,0)(0,-45){-2}{4}
\Text(0,-50)[l]{$q$}
\PhotonArc(0,0)(40,0,180){4}{7.5}
\Text(0,50)[c]{$W(l - p)$}
\end{picture}
\quad\quad
% graviton coupled to W
%
\begin{picture}(160,130)(-80,-65)
\SetWidth{1.2}
\Text(0,-35)[ct]{(b)}
\ArrowLine(80,0)(40,0)
\Text(60,-10)[c]{$\nu_e(p)$}
\ArrowLine(40,0)(-40,0)
\Text(0,-10)[c]{$e(l)$}
\ArrowLine(-40,0)(-80,0)
\Text(-60,-10)[cr]{$\nu_e(p^\prime)$}
\Photon(0,44)(0,80){2}{4}
\Photon(0,44)(0,80){-2}{4}
\Text(0,85)[bl]{$q$}
\PhotonArc(0,0)(40,0,180){4}{6.5}
\Text(30,40)[l]{$W(l - p)$}
\Text(-30,40)[r]{$W(l - p^\prime)$}
\end{picture}
%
% Contact term - right
%
\begin{picture}(160,130)(-80,-65)
\SetWidth{1.2}
\Text(0,-35)[ct]{(c)}
\ArrowLine(80,0)(40,0)
\Text(60,-10)[c]{$\nu_e(p)$}
\ArrowLine(40,0)(-40,0)
\Text(0,-10)[c]{$e(l)$}
\ArrowLine(-40,0)(-80,0)
\Text(-60,-10)[c]{$\nu_e(p^\prime)$}
\PhotonArc(0,0)(40,0,180){4}{7.5}
\Text(0,50)[c]{$W(l - p^\prime)$}
\Photon(40,0)(40,-50){2}{4}
\Photon(40,0)(40,-50){-2}{4}
\Text(40,-55)[l]{$q$}
\end{picture}
\quad\quad
% Contact term - left
%
\begin{picture}(160,130)(-80,-65)
\SetWidth{1.2}
\Text(0,-35)[ct]{(d)}
\ArrowLine(80,0)(40,0)
\Text(60,-10)[c]{$\nu_e(p)$}
\ArrowLine(40,0)(-40,0)
\Text(0,-10)[c]{$e(l)$}
\ArrowLine(-40,0)(-80,0)
\Text(-60,-10)[c]{$\nu_e(p^\prime)$}
\PhotonArc(0,0)(40,0,180){4}{7.5}
\Text(0,50)[c]{$W(l - p)$}
\Photon(-40,0)(-40,-50){2}{4}
\Photon(-40,0)(-40,-50){-2}{4}
\Text(-40,-55)[r]{$q$}
\end{picture}
\end{center}
\caption[]{$W$ exchange diagrams for the one-loop contribution
to the $\nu_e$ gravitational vertex in a background of electrons. The
braided line represents the graviton.
\label{s2:Wdiagrams}}
\end{figure}
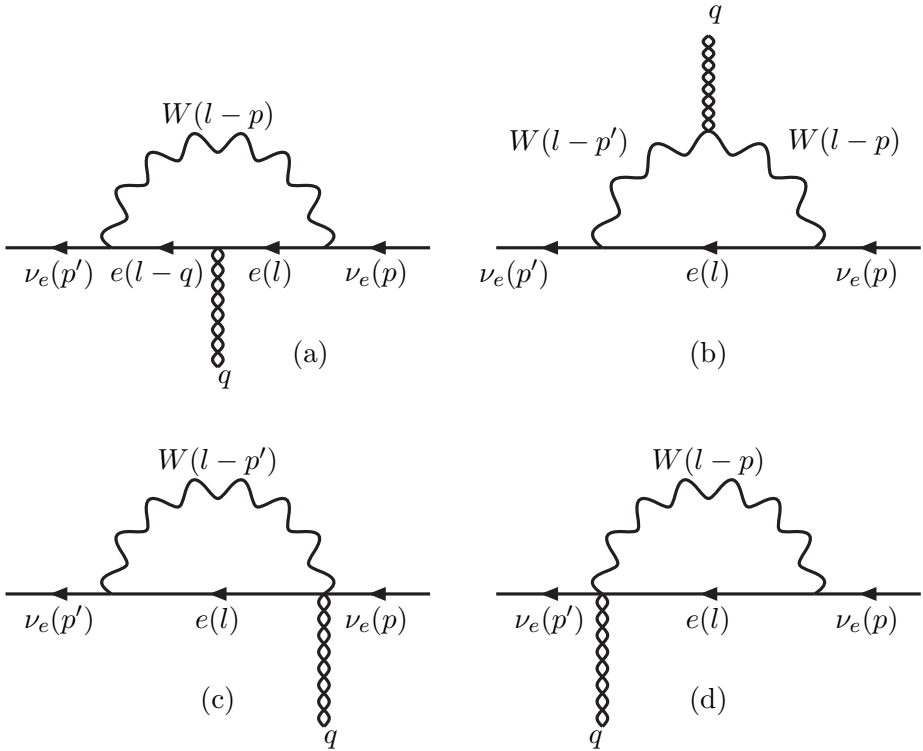

We take the medium to be a thermal background of electrons.
To one-loop order, $\Gamma_{\lambda\rho}$ is determined by the
diagrams shown in \Fig{s2:Wdiagrams}.
Those diagrams contribute only to the gravitational coupling of
$\nu_e$ and not to the other flavors, which lead to non-diagonal
couplings between the neutrino mass eigenstates through the
neutrino mixing matrix. There is another set of
one-loop diagrams involving the $Z$ boson, but they
contribute equally to the gravitational vertex of all the neutrino
flavors $\nu_{e,\mu,\tau}$, and therefore do not contribute to the
off-diagonal couplings between the mass eigenstates of
standard, weak SU(2)-doublet neutrinos.
They are relevant if the decay process involves a so-called
sterile neutrino, but we do not consider that possibility here.

There are additional diagrams in which one or both of the
vector boson lines are replaced by their corresponding unphysical
Higgs partners. As we will see, the leading contributions from the $W$
exchange diagrams to the decay amplitude is $O(1/M^2_W)$, and
therefore the Higgs exchange diagrams are relevant in
general. However, for simplicity, we will carry out the calculation in
the limit $m_e\rightarrow 0$, in which case the diagrams involving the
unphysical Higgs particles do not contribute. Thus, our results are
strictly applicable for those physical environments in which the
electrons in the background can be considered to be relativistic.

There are also diagrams in which the graviton line comes out from one
of the external neutrino legs are one-particle reducible and do not
contribute to the leading order term of the off-diagonal vertex
function $\Gamma_{\lambda\rho}$.  The proper way to take those
diagrams into account in the calculation of the amplitude for any
given process, is by choosing the external neutrino spinors $u_i(p)$
to be the (properly normalized) solutions of the effective Dirac
equation for the propagating neutrino mode in the medium, instead of
the spinor representing the free-particle solution of the equation in
the vacuum.  Since the off-diagonal vertex function is zero at the
tree-level, those corrections yield higher order contributions to the
decay amplitude, which we neglect.

Therefore, we will compute the integral expressions defined by the
diagrams shown in \Fig{s2:Wdiagrams}, in a manner that is consistent
with the approximations and idealizations that we have outlined above.

%%%
\subsection{One-loop expressions for the vertex} 
%%%
We consider situations in which the $W$-boson mass is much larger than
all the other energy scales, so that it is valid to neglect the
thermal effects in the $W$-boson propagator. In addition, we can
expand the amplitude in inverse powers of the $W$-boson mass and keep
only the leading terms. The one-loop
expressions for the neutrino gravitational vertex function under the
above conditions has been determined in previous
works \cite{Nieves:1998xz, Nieves:1999rt,Nieves:2000dc},
and the results that are relevant for our present purposes
can be summarized as follows.

Up to terms of order $1/M^4_W$, the vertex function can be written in
the form
\begin{eqnarray}
\label{s2:oneloopgamma}
\Gamma_{\lambda\rho} = U_{e1} U_{e2}^* \left[\Lambda_{\lambda\rho} +
G_{\lambda\rho} + H_{\lambda\rho}\right]\,,
\end{eqnarray}
where $U$ is the neutrino mixing matrix.
The terms contained in $\Lambda_{\lambda\rho}$
are the leading order terms in the Fermi constant
and they were derived in the first two references cited above. Those terms
do not contribute to the on-shell amplitude, which is most easily seen
by noticing that those terms are momentum-independent and therefore do not
produce any contribution of the type shown in \Eq{chiralamplitudeonshell}.

The remaining terms in \Eq{s2:oneloopgamma} are the $O(1/M^4_W)$
terms, which were determined in \Ref{Nieves:2000dc}.  They are
momentum-dependent and the relevant ones for the present work.  The
part denoted $G_{\lambda\rho}$ contains all the $O(1/M^4_W)$ terms
that are independent of the weak gauge parameter $\xi$, while
$H_{\lambda\rho}$ contains the rest, which depend on $\xi$.  Borrowing
the results from that reference, we quote below the relevant formulas
for the contribution from each of the diagrams in \Fig{s2:Wdiagrams}
to these two sets of terms.

Let us consider $G_{\lambda\rho}$ first, which contains
all the $O(1/M^4_W)$ terms that are $\xi$-independent, and
let us start with diagram (a). The contribution from this diagram
is given by
\begin{eqnarray}
\label{Ga}
G^{(a)}_{\lambda\rho} & = & -\frac{g^2}{2}
\int\frac{d^4 l}{(2\pi)^3}\delta(l^2) \eta(l)
\gamma^\alpha L \left[
\frac{(\lslash{l} - \lslash{q})
V_{\lambda\rho}(l,l - q)\lslash{l}
\Delta^{(4)}_{W\alpha\beta}(l - p)}{(l - q)^2}\right.\nonumber\\*
&&\mbox{} + \left.\frac{\lslash{l}
V_{\lambda\rho}(l + q,l)(\lslash{l} + \lslash{q})
\Delta^{(4)}_{W\alpha\beta}(l - p^\prime)}{(l + q)^2}
\right]\gamma^\beta L\,,
\end{eqnarray}
where
\begin{eqnarray}
\label{Delta4}
\Delta^{(4)}_{W\mu\nu}(k) = \frac{1}{M_W^4}\left(
k^2\eta_{\mu\nu} - k_\mu k_\nu\right) \,,
\end{eqnarray}
\begin{eqnarray}
V_{\lambda\rho}(k,k') = \frac14 \Big[ \gamma_\lambda (k+k')_\rho +
\gamma_\rho (k+k')_\lambda \Big] - \frac12\eta_{\lambda\rho}
\Big[ \rlap/k + \rlap/k'\Big]
\end{eqnarray}
is the tree-level electron gravitational coupling (in the massless limit),
and $\eta(l)$ has been defined in \Eq{etae}.
The factor of $\delta(l^2)$ appears in \Eq{Ga} because
we have taken the limit $m_e\rightarrow 0$, as explained earlier.  It should be
noted that we have retained only the contribution due to the
thermal part of the electron propagator since,
as already mentioned, the on-shell amplitude vanishes in the vacuum.
Similarly, for the other diagrams,
\begin{eqnarray}
\label{Gbcd}
G^{(b)}_{\lambda\rho} &=& \frac{g^2}{2}
\int\frac{d^4l}{(2\pi)^3}\delta(l^2)\eta(l)
\gamma^\alpha L\rlap/l 
\gamma^\beta L \nonumber\\* 
&\times& 
\left[-\frac{1}{M_W^4}C_{\lambda\rho\alpha\beta}(l - p^\prime,l - p)  
+ \eta^{\mu\nu} a^\prime_{\lambda\rho\alpha\mu}
\Delta^{(4)}_{W\nu\beta}(l - p) 
+ \eta^{\mu\nu} a^\prime_{\lambda\rho\mu\beta}
\Delta^{(4)}_{W\alpha\nu}(l - p^\prime)\right] \,,\nonumber\\
G^{(c)}_{\lambda\rho} &=& -\, {g^2 \over 2} 
a_{\lambda\rho\beta\mu}
\int\frac{d^4l}{(2\pi)^3}\delta(l^2)\eta(p)
\gamma_\alpha L \rlap/l \gamma^\beta L
\Delta^{(4)\mu\alpha}_W(l - p')\,,\nonumber\\
G^{(d)}_{\lambda\rho} &=& -\, {g^2 \over 2} 
a_{\lambda\rho\alpha\mu}
\int\frac{d^4l}{(2\pi)^3}\delta(l^2)\eta(l)
\gamma^\alpha L \rlap/l \gamma_\beta L 
\Delta^{(4)\beta\mu}_W(l - p)\,,
\end{eqnarray}
where the tensors $a_{\lambda\rho\mu\nu}$, $a^\prime_{\lambda\rho\mu\nu}$,
and $C_{\lambda\rho\mu\nu}$ that appear in these expressions are
related to the various gravitational vertices that appear in the diagrams,
and are given by
\begin{eqnarray}
a_{\lambda\rho\mu\nu} 
& = & \eta_{\lambda\rho}\eta_{\mu\nu} - \frac{1}{2}(
\eta_{\lambda\mu}\eta_{\rho\nu} +
\eta_{\rho\mu}\eta_{\lambda\nu}) \,,\nonumber\\
a^\prime_{\lambda\rho\mu\nu} & = & \eta_{\lambda\rho} \eta_{\mu\nu} -
\big( \eta_{\lambda\mu} \eta_{\rho\nu} +
\eta_{\lambda\nu} \eta_{\rho\mu} \big) \,,\nonumber\\
C_{\lambda\rho\mu\nu} (k,k') &=& \eta_{\lambda\rho} ( \eta_{\mu\nu} k
\cdot k' - k^\prime_\mu k_\nu ) - \eta_{\mu\nu} (k_\lambda k'_\rho +
k'_\lambda k_\rho ) \nonumber\\*
&& \mbox{} + k_\nu (\eta_{\lambda\mu} k'_\rho +
\eta_{\rho\mu} k'_\lambda)
+ k'_\mu (\eta_{\lambda\nu} k_\rho +
\eta_{\rho\nu} k_\lambda) \nonumber\\*
&& \mbox{} - k \cdot k' (\eta_{\lambda\mu} \eta_{\rho\nu} +
\eta_{\lambda\nu} \eta_{\rho\mu})\,.
\end{eqnarray}

The expression for the $H^{(a,b,c,d)}_{\lambda\rho}$, which
contain the $\xi$-dependent terms, are given by the same expressions
as \Eqs{Ga}{Gbcd}, but with the replacements
\begin{eqnarray}
\label{replacement} 
\Delta^{(4)}_{W\alpha\beta} & \rightarrow &
\Delta^{(\xi)}_{W\alpha\beta} = \frac{\xi k_\mu k_\nu}{M_W^4} \,,
\nonumber\\*
C_{\lambda\rho\mu\nu} (k,k') & \rightarrow & 
\xi\left\{
\eta_{\lambda\rho}k_\mu k^\prime_\nu - k^\prime_\lambda k_\mu\eta_{\rho\nu}
- k^\prime_\rho k_\mu\eta_{\lambda\nu} - k_\lambda k^\prime_\nu\eta_{\rho\mu}
- k_\rho k^\prime_\nu \eta_{\lambda\mu}\right\} \,.
\end{eqnarray}

Before extracting the physical part from these expressions, we consider
the conditions required by the gauge invariance of the
weak and the gravitational interactions.

\subsection{Weak gauge invariance}
The gauge invariance of the weak interactions requires that the on-shell
amplitude be independent of the parameter $\xi$. It turns out that
the $H$ terms satisfy
\begin{eqnarray}
\label{weakgaugeinv}
\bar u_2(p^\prime) H^{(x)}_{\lambda\rho} u_1(p) = 0 \qquad (x = a,b,c,d)\,,
\end{eqnarray}
so that the requirement is indeed satisfied. In order to show this,
let us consider specifically the first term $x = a$, where
\begin{eqnarray}
H^{(a)}_{\lambda\rho} & = & -\frac{g^2}{2}
\int\frac{d^4 l}{(2\pi)^3}\delta(l^2) \eta(l)
\gamma^\alpha L \left[
\frac{(\lslash{l} - \lslash{q})
V_{\lambda\rho}(l,l - q)\lslash{l}
\Delta^{(\xi)}_{W\alpha\beta}(l - p)}{(l -
q)^2}\right.\nonumber\\* 
&&\mbox{} + \left.\frac{\lslash{l}
V_{\lambda\rho}(l + q,l)(\lslash{l} + \lslash{q})
\Delta^{(\xi)}_{W\alpha\beta}(l - p^\prime)}{(l + q)^2}
\right]\gamma^\beta L\,.
\end{eqnarray}
Using \Eq{replacement}, the first term in the square brackets involves
the factor
$
\lslash{l}(\lslash{l} - \lslash{p})u_1(p)\,,
$
which reduces to zero
when the massless Dirac equation for $u_1$ and the delta function
$\delta(l^2)$ are taken into account. Similarly, the second term
in square brackets contains the factor
$
\bar u_2(p^\prime)
(\lslash{l} - \lslash{p^\prime})\lslash{l}\,,
$
which also reduces to zero for similar reasons.

Similar arguments hold for the remaining $H$-terms, which we do not
consider explicitly any further. Therefore, the conclusion is
that only the $G$ terms, which are independent of $\xi$,
contribute to the physical amplitude. 

\subsection{Physical part of the amplitude}
A by-product of \Eq{weakgaugeinv} is that the $k_\mu k_\nu$ term
in \Eq{Delta4} does not contribute to the physical amplitude
and therefore for the present purposes we can adopt
\begin{eqnarray}
\Delta^{(4)}_{W\mu\nu}(k) = \frac{k^2}{M_W^4}\eta_{\mu\nu}\,,
\end{eqnarray}
in \Eqs{Ga}{Gbcd}. In the evaluation of the expressions the
$G^{(x)}_{\lambda\rho}$,
we can ignore the terms that do not contribute to the on-shell
amplitude due to the on-shell relations given in \Eqs{gravonshell}{nuonshell}.
Denoting the latter by $G^{\,\prime\prime\,(x)}_{\lambda\rho}$
we then write
\begin{eqnarray}
\label{physicalG}
G^{(x)}_{\lambda\rho} = \left(\frac{g^2}{M^4_W}\right)
G^{\,\prime\,(x)}_{\lambda\rho} + G^{\,\prime\prime\,(x)}_{\lambda\rho}\,,
\end{eqnarray}
where the
$G^{\,\prime\,(x)}_{\lambda\rho}$ contain the terms that survive.
A straightforward evaluation of the integral expressions then yields
\begin{eqnarray}
G_{\lambda\rho}^{\,\prime\,(a)} &=& \left[
2J_{\lambda\rho\alpha} + p_\lambda J_{\rho\alpha} + p_\rho J_{\lambda\alpha}
\right]\gamma^\alpha L\,,\nonumber \\ 
G_{\lambda\rho}^{\,\prime\,(b)} &=& -2\Big[ J_{\alpha\lambda\rho}
  - p_\lambda J_{\alpha\rho} - p_\rho J_{\alpha\lambda} + p_\lambda
  p_\rho J_\alpha \Big]\gamma^\alpha L \nonumber \\*
&& + \left[J_{\lambda\alpha} - p_\lambda J_\alpha\right]\left[
(p + p^\prime)^\alpha\gamma_\rho - 2 p_\rho\gamma^\alpha \right]L\nonumber\\*
&&
+ \left[J_{\rho\alpha} - p_\rho J_\alpha\right]\left[
(p + p^\prime)^\alpha\gamma_\lambda - 2 p_\lambda\gamma^\alpha
\right]L\nonumber\\*
&& + \Big[J^{\alpha\beta}(p + p')_\beta - p \cdot p'
  J^\alpha \Big](\eta_{\lambda\alpha}\gamma_\rho + \eta_{\rho\alpha}
\gamma_\lambda)L \,, \nonumber \\
G_{\lambda\rho}^{\,\prime\,(c)} &=& - p'^\alpha \Big(
J_{\lambda\alpha}\gamma_\rho + J_{\rho\alpha}\gamma_\lambda  \Big)L
\,, \nonumber \\ 
G_{\lambda\rho}^{\,\prime\,(d)} &=& - p^\alpha \Big(
J_{\lambda\alpha}\gamma_\rho + J_{\rho\alpha}\gamma_\lambda  \Big)L \,,
\label{Gabcd}
\end{eqnarray}
where we have introduced the definitions
\begin{eqnarray}
J_{\alpha_1\alpha_2\cdots\alpha_n} = \int {d^4l \over (2\pi)^3} \;
\delta(l^2) \eta(l) \; l_{\alpha_1} l_{\alpha_2} \cdots l_{\alpha_n}\,.
\label{defJ}
\end{eqnarray}

The integral $J_{\alpha\lambda\rho}$ cancels when all the terms are added,
while the Lorentz structure of $J_\alpha$ and $J_{\alpha\beta}$
imply that they are of the form
\begin{eqnarray}
\label{Adef}
J_\alpha &=& A_{11} v_\alpha \,, \nonumber\\*
J_{\alpha\beta} &=& A_{20} \eta_{\alpha\beta} + A_{22} v_\alpha
v_\beta \,.
\end{eqnarray}
Thus, substituting \Eq{Adef} into \Eq{Gabcd} we obtain
\begin{eqnarray}
\label{Gfinal}
G^\prime_{\lambda\rho} = 2A_{11} p_\lambda p_\rho\lslash{v}L +
[3A_{20} - A_{11}(p+p') \cdot v ]
(p_\mu \gamma_\rho + p_\rho\gamma_\lambda)L \,.
\end{eqnarray}
which, together with \Eqs{s2:oneloopgamma}{physicalG},
establishes that the amplitude is of the form
given in \Eq{chiralamplitudeonshell}, with 
\begin{eqnarray}
\label{formfactorsresult}
c &=& 2U_{e1} U_{e2}^* \; {g^2 \over M_W^4} A_{11} \; \nonumber\\*
d &=& U_{e1} U_{e2}^* \; {g^2 \over M_W^4} [3A_{20} - (2E - \omega)A_{11}] \,.
%\label{}
\end{eqnarray}

\subsection{Evaluation of the form factors}
The definitions in \Eq{Adef} imply the following expressions for
the coefficients $A_{11}$ and $A_{20}$,
\begin{eqnarray}
A_{11} & = & v^\lambda J_\lambda, \nonumber\\*
A_{20} & = & \frac{1}{3}\left({J^\lambda}_\lambda -
v^\lambda v^\rho J_{\lambda\rho}\right)\,,
\end{eqnarray}
which give the integral formulas
\begin{eqnarray}
A_{11} & = & \frac{1}{2}\int\frac{d^3l}{(2\pi)^3}
\left(f_e - f_{\bar e}\right) \,,\nonumber\\
A_{20} & = & -\frac{1}{6}\int\frac{d^3l}{(2\pi)^3} \; |\vec l| 
\left(f_e + f_{\bar e}\right)
\end{eqnarray}
where $f_{e,\bar e}$ are the distribution functions of electrons and
positrons respectively. In the massless limit that we have employed
these integrals can be evaluated exactly in terms of the temperature $T$ and
chemical potential $\mu$ of the electron gas, to yield
\begin{eqnarray}
\label{n&rho}
A_{11} &=& {\mu \over 12}\left(T^2 + {\mu^2 \over \pi^2}\right) \,,
\nonumber\\
A_{20} &=& -\frac{1}{12}\left[{7\pi^2 \over 60} T^4 + \frac12 \mu^2 T^2 +
{1 \over 4\pi^2} \; \mu^4\right] \,.
\end{eqnarray}
The explicit formulas for the form factors $c$ and $d$
are obtained by substituting \Eq{n&rho} in \Eq{formfactorsresult}.

%%%%%
\subsection{Decay rate}
\label{sec:decayrate}
%%%%%

Substituting \Eq{formfactorsresult} in \Eq{Msquared},
\begin{eqnarray}
\label{Msquaredresult}
\overline{|{\cal M}|^2} = 8\left|U_{e1}U_{e2}^*\right|^2
\left(\frac{g^2 \kappa}{M_W^4}\right)^2
P^4 \sin^4\theta\left(C_1 + C_2 \frac{\omega}{\omega_0}\right)\,,
\end{eqnarray}
where
\begin{eqnarray}
C_1 & = & (3A_{20} - EA_{11})^2 \,, \nonumber\\*
C_2 & = & \omega_0 A_{11}(3A_{20} - EA_{11})  \,,
\label{C1C2}
\end{eqnarray}
with $\omega_0$ given in \Eq{omega0}. The decay rate is then
obtained by substituting \Eq{Msquaredresult} into \Eq{totalrateformula}.  Using
\Eq{gravitonomega} to express $\theta$ in terms of $\omega$ and making
the change of variable $\omega = \omega_0 y$, we get
\begin{eqnarray}
\Gamma = \frac{\omega_0}{2\pi V} 
\left( \frac{g^2 \kappa m}{M_W^4} |U_{e1}U_{e2}^*|
\right)^2 I\,, 
\label{defI}
\end{eqnarray}
where
\begin{eqnarray}
I = \int_{r}^{1/r} dy
\left[ \sqrt{1-V^2} \left( 1+ {1 \over y^2} \right) - {2\over y}
\right]^2 \left(C_1 + C_2 y\right)\,,
\end{eqnarray}
The integral $I$ is of course trivial but the resulting formulas are
cumbersome. Introducing
\begin{eqnarray}
R_n \equiv \int^{1/r}_r dy\,\frac{1}{y^n} = \left\{
\begin{array}{ll}
\log\left(\frac{1}{r^2}\right) & \mbox{$n = 1$} \\[12pt]
\frac{1}{n - 1}\left[r^{-n + 1} - r^{n - 1}\right] & \mbox{$n \not= 1$}\,,
\end{array}
\right.
\end{eqnarray}
and noting that $R_{-n+2} = R_n$, the result can be written in the form
\begin{eqnarray}
I = \sum_{n = 1}^4 \alpha_n R_n\,,
\end{eqnarray}
where
\begin{eqnarray}
\alpha_1 & = &  - 4C_1 \sqrt{1-V^2} + 2C_2 (3-V^2) \,, \nonumber\\
\alpha_2 & = & C_1 (7 - 3V^2) - 8C_2 \sqrt{1-V^2} \,, \nonumber\\
\alpha_3 & = & - 4C_1 \sqrt{1-V^2} + 2C_2 (1-V^2) \,, \nonumber\\*
\alpha_4 & = & C_1 (1-V^2) \,.
\end{eqnarray}
It should be noted that, for small $V$, 
\begin{eqnarray}
R_n = 2V + O(V^2)
%\label{}
\end{eqnarray}
and 
\begin{eqnarray}
\sum_n \alpha_n = O(V^2) \,,
%\label{}
\end{eqnarray}
which imply that
\begin{eqnarray}
\sum_n \alpha_n R_n = O(V^2) \,. 
%\label{}
\end{eqnarray}
Thus, despite the overall factor of $1/V$ in \Eq{defI}, the rate
vanishes for $V=0$, which confirms the statements given in the Introduction
based on general grounds.

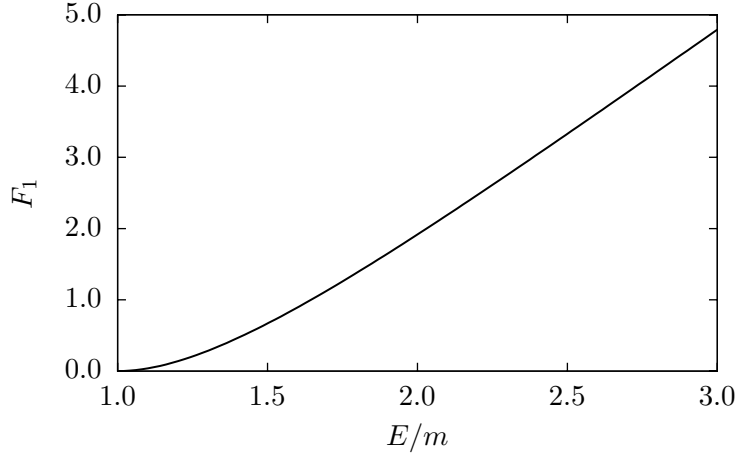
\begin{figure}
\begin{center}
% GNUPLOT: LaTeX picture with Postscript
\begingroup%
\makeatletter%
\newcommand{\GNUPLOTspecial}{%
  \@sanitize\catcode`\%=14\relax\special}%
\setlength{\unitlength}{0.0500bp}%
\begin{picture}(5760,3527)(0,0)%
  {\GNUPLOTspecial{"
%!PS-Adobe-2.0 EPSF-2.0
%%Title: amvdk_fig3.psl
%%Creator: gnuplot 4.2 patchlevel 0
%%CreationDate: Thu Jun  4 14:19:53 2009
%%DocumentFonts: 
%%BoundingBox: 0 0 288 176
%%EndComments
%%BeginProlog
/gnudict 256 dict def
gnudict begin
%
% The following 6 true/false flags may be edited by hand if required
% The unit line width may also be changed
%
/Color false def
/Blacktext false def
/Solid false def
/Dashlength 1 def
/Landscape false def
/Level1 false def
/Rounded false def
/TransparentPatterns false def
/gnulinewidth 5.000 def
/userlinewidth gnulinewidth def
/vshift -66 def
/dl1 {
  10.0 Dashlength mul mul
  Rounded { currentlinewidth 0.75 mul sub dup 0 le { pop 0.01 } if } if
} def
/dl2 {
  10.0 Dashlength mul mul
  Rounded { currentlinewidth 0.75 mul add } if
} def
/hpt_ 31.5 def
/vpt_ 31.5 def
/hpt hpt_ def
/vpt vpt_ def
Level1 {} {
/SDict 10 dict def
systemdict /pdfmark known not {
  userdict /pdfmark systemdict /cleartomark get put
} if
SDict begin [
  /Title (amvdk_fig3.psl)
  /Subject (gnuplot plot)
  /Creator (gnuplot 4.2 patchlevel 0)
  /Author (Palash Baran Pal)
%  /Producer (gnuplot)
%  /Keywords ()
  /CreationDate (Thu Jun  4 14:19:53 2009)
  /DOCINFO pdfmark
end
} ifelse
%
% Gnuplot Prolog Version 4.2 (August 2006)
%
/M {moveto} bind def
/L {lineto} bind def
/R {rmoveto} bind def
/V {rlineto} bind def
/N {newpath moveto} bind def
/Z {closepath} bind def
/C {setrgbcolor} bind def
/f {rlineto fill} bind def
/vpt2 vpt 2 mul def
/hpt2 hpt 2 mul def
/Lshow {currentpoint stroke M 0 vshift R 
	Blacktext {gsave 0 setgray show grestore} {show} ifelse} def
/Rshow {currentpoint stroke M dup stringwidth pop neg vshift R
	Blacktext {gsave 0 setgray show grestore} {show} ifelse} def
/Cshow {currentpoint stroke M dup stringwidth pop -2 div vshift R 
	Blacktext {gsave 0 setgray show grestore} {show} ifelse} def
/UP {dup vpt_ mul /vpt exch def hpt_ mul /hpt exch def
  /hpt2 hpt 2 mul def /vpt2 vpt 2 mul def} def
/DL {Color {setrgbcolor Solid {pop []} if 0 setdash}
 {pop pop pop 0 setgray Solid {pop []} if 0 setdash} ifelse} def
/BL {stroke userlinewidth 2 mul setlinewidth
	Rounded {1 setlinejoin 1 setlinecap} if} def
/AL {stroke userlinewidth 2 div setlinewidth
	Rounded {1 setlinejoin 1 setlinecap} if} def
/UL {dup gnulinewidth mul /userlinewidth exch def
	dup 1 lt {pop 1} if 10 mul /udl exch def} def
/PL {stroke userlinewidth setlinewidth
	Rounded {1 setlinejoin 1 setlinecap} if} def
% Default Line colors
/LCw {1 1 1} def
/LCb {0 0 0} def
/LCa {0 0 0} def
/LC0 {1 0 0} def
/LC1 {0 1 0} def
/LC2 {0 0 1} def
/LC3 {1 0 1} def
/LC4 {0 1 1} def
/LC5 {1 1 0} def
/LC6 {0 0 0} def
/LC7 {1 0.3 0} def
/LC8 {0.5 0.5 0.5} def
% Default Line Types
/LTw {PL [] 1 setgray} def
/LTb {BL [] LCb DL} def
/LTa {AL [1 udl mul 2 udl mul] 0 setdash LCa setrgbcolor} def
/LT0 {PL [] LC0 DL} def
/LT1 {PL [4 dl1 2 dl2] LC1 DL} def
/LT2 {PL [2 dl1 3 dl2] LC2 DL} def
/LT3 {PL [1 dl1 1.5 dl2] LC3 DL} def
/LT4 {PL [6 dl1 2 dl2 1 dl1 2 dl2] LC4 DL} def
/LT5 {PL [3 dl1 3 dl2 1 dl1 3 dl2] LC5 DL} def
/LT6 {PL [2 dl1 2 dl2 2 dl1 6 dl2] LC6 DL} def
/LT7 {PL [1 dl1 2 dl2 6 dl1 2 dl2 1 dl1 2 dl2] LC7 DL} def
/LT8 {PL [2 dl1 2 dl2 2 dl1 2 dl2 2 dl1 2 dl2 2 dl1 4 dl2] LC8 DL} def
/Pnt {stroke [] 0 setdash gsave 1 setlinecap M 0 0 V stroke grestore} def
/Dia {stroke [] 0 setdash 2 copy vpt add M
  hpt neg vpt neg V hpt vpt neg V
  hpt vpt V hpt neg vpt V closepath stroke
  Pnt} def
/Pls {stroke [] 0 setdash vpt sub M 0 vpt2 V
  currentpoint stroke M
  hpt neg vpt neg R hpt2 0 V stroke
 } def
/Box {stroke [] 0 setdash 2 copy exch hpt sub exch vpt add M
  0 vpt2 neg V hpt2 0 V 0 vpt2 V
  hpt2 neg 0 V closepath stroke
  Pnt} def
/Crs {stroke [] 0 setdash exch hpt sub exch vpt add M
  hpt2 vpt2 neg V currentpoint stroke M
  hpt2 neg 0 R hpt2 vpt2 V stroke} def
/TriU {stroke [] 0 setdash 2 copy vpt 1.12 mul add M
  hpt neg vpt -1.62 mul V
  hpt 2 mul 0 V
  hpt neg vpt 1.62 mul V closepath stroke
  Pnt} def
/Star {2 copy Pls Crs} def
/BoxF {stroke [] 0 setdash exch hpt sub exch vpt add M
  0 vpt2 neg V hpt2 0 V 0 vpt2 V
  hpt2 neg 0 V closepath fill} def
/TriUF {stroke [] 0 setdash vpt 1.12 mul add M
  hpt neg vpt -1.62 mul V
  hpt 2 mul 0 V
  hpt neg vpt 1.62 mul V closepath fill} def
/TriD {stroke [] 0 setdash 2 copy vpt 1.12 mul sub M
  hpt neg vpt 1.62 mul V
  hpt 2 mul 0 V
  hpt neg vpt -1.62 mul V closepath stroke
  Pnt} def
/TriDF {stroke [] 0 setdash vpt 1.12 mul sub M
  hpt neg vpt 1.62 mul V
  hpt 2 mul 0 V
  hpt neg vpt -1.62 mul V closepath fill} def
/DiaF {stroke [] 0 setdash vpt add M
  hpt neg vpt neg V hpt vpt neg V
  hpt vpt V hpt neg vpt V closepath fill} def
/Pent {stroke [] 0 setdash 2 copy gsave
  translate 0 hpt M 4 {72 rotate 0 hpt L} repeat
  closepath stroke grestore Pnt} def
/PentF {stroke [] 0 setdash gsave
  translate 0 hpt M 4 {72 rotate 0 hpt L} repeat
  closepath fill grestore} def
/Circle {stroke [] 0 setdash 2 copy
  hpt 0 360 arc stroke Pnt} def
/CircleF {stroke [] 0 setdash hpt 0 360 arc fill} def
/C0 {BL [] 0 setdash 2 copy moveto vpt 90 450 arc} bind def
/C1 {BL [] 0 setdash 2 copy moveto
	2 copy vpt 0 90 arc closepath fill
	vpt 0 360 arc closepath} bind def
/C2 {BL [] 0 setdash 2 copy moveto
	2 copy vpt 90 180 arc closepath fill
	vpt 0 360 arc closepath} bind def
/C3 {BL [] 0 setdash 2 copy moveto
	2 copy vpt 0 180 arc closepath fill
	vpt 0 360 arc closepath} bind def
/C4 {BL [] 0 setdash 2 copy moveto
	2 copy vpt 180 270 arc closepath fill
	vpt 0 360 arc closepath} bind def
/C5 {BL [] 0 setdash 2 copy moveto
	2 copy vpt 0 90 arc
	2 copy moveto
	2 copy vpt 180 270 arc closepath fill
	vpt 0 360 arc} bind def
/C6 {BL [] 0 setdash 2 copy moveto
	2 copy vpt 90 270 arc closepath fill
	vpt 0 360 arc closepath} bind def
/C7 {BL [] 0 setdash 2 copy moveto
	2 copy vpt 0 270 arc closepath fill
	vpt 0 360 arc closepath} bind def
/C8 {BL [] 0 setdash 2 copy moveto
	2 copy vpt 270 360 arc closepath fill
	vpt 0 360 arc closepath} bind def
/C9 {BL [] 0 setdash 2 copy moveto
	2 copy vpt 270 450 arc closepath fill
	vpt 0 360 arc closepath} bind def
/C10 {BL [] 0 setdash 2 copy 2 copy moveto vpt 270 360 arc closepath fill
	2 copy moveto
	2 copy vpt 90 180 arc closepath fill
	vpt 0 360 arc closepath} bind def
/C11 {BL [] 0 setdash 2 copy moveto
	2 copy vpt 0 180 arc closepath fill
	2 copy moveto
	2 copy vpt 270 360 arc closepath fill
	vpt 0 360 arc closepath} bind def
/C12 {BL [] 0 setdash 2 copy moveto
	2 copy vpt 180 360 arc closepath fill
	vpt 0 360 arc closepath} bind def
/C13 {BL [] 0 setdash 2 copy moveto
	2 copy vpt 0 90 arc closepath fill
	2 copy moveto
	2 copy vpt 180 360 arc closepath fill
	vpt 0 360 arc closepath} bind def
/C14 {BL [] 0 setdash 2 copy moveto
	2 copy vpt 90 360 arc closepath fill
	vpt 0 360 arc} bind def
/C15 {BL [] 0 setdash 2 copy vpt 0 360 arc closepath fill
	vpt 0 360 arc closepath} bind def
/Rec {newpath 4 2 roll moveto 1 index 0 rlineto 0 exch rlineto
	neg 0 rlineto closepath} bind def
/Square {dup Rec} bind def
/Bsquare {vpt sub exch vpt sub exch vpt2 Square} bind def
/S0 {BL [] 0 setdash 2 copy moveto 0 vpt rlineto BL Bsquare} bind def
/S1 {BL [] 0 setdash 2 copy vpt Square fill Bsquare} bind def
/S2 {BL [] 0 setdash 2 copy exch vpt sub exch vpt Square fill Bsquare} bind def
/S3 {BL [] 0 setdash 2 copy exch vpt sub exch vpt2 vpt Rec fill Bsquare} bind def
/S4 {BL [] 0 setdash 2 copy exch vpt sub exch vpt sub vpt Square fill Bsquare} bind def
/S5 {BL [] 0 setdash 2 copy 2 copy vpt Square fill
	exch vpt sub exch vpt sub vpt Square fill Bsquare} bind def
/S6 {BL [] 0 setdash 2 copy exch vpt sub exch vpt sub vpt vpt2 Rec fill Bsquare} bind def
/S7 {BL [] 0 setdash 2 copy exch vpt sub exch vpt sub vpt vpt2 Rec fill
	2 copy vpt Square fill Bsquare} bind def
/S8 {BL [] 0 setdash 2 copy vpt sub vpt Square fill Bsquare} bind def
/S9 {BL [] 0 setdash 2 copy vpt sub vpt vpt2 Rec fill Bsquare} bind def
/S10 {BL [] 0 setdash 2 copy vpt sub vpt Square fill 2 copy exch vpt sub exch vpt Square fill
	Bsquare} bind def
/S11 {BL [] 0 setdash 2 copy vpt sub vpt Square fill 2 copy exch vpt sub exch vpt2 vpt Rec fill
	Bsquare} bind def
/S12 {BL [] 0 setdash 2 copy exch vpt sub exch vpt sub vpt2 vpt Rec fill Bsquare} bind def
/S13 {BL [] 0 setdash 2 copy exch vpt sub exch vpt sub vpt2 vpt Rec fill
	2 copy vpt Square fill Bsquare} bind def
/S14 {BL [] 0 setdash 2 copy exch vpt sub exch vpt sub vpt2 vpt Rec fill
	2 copy exch vpt sub exch vpt Square fill Bsquare} bind def
/S15 {BL [] 0 setdash 2 copy Bsquare fill Bsquare} bind def
/D0 {gsave translate 45 rotate 0 0 S0 stroke grestore} bind def
/D1 {gsave translate 45 rotate 0 0 S1 stroke grestore} bind def
/D2 {gsave translate 45 rotate 0 0 S2 stroke grestore} bind def
/D3 {gsave translate 45 rotate 0 0 S3 stroke grestore} bind def
/D4 {gsave translate 45 rotate 0 0 S4 stroke grestore} bind def
/D5 {gsave translate 45 rotate 0 0 S5 stroke grestore} bind def
/D6 {gsave translate 45 rotate 0 0 S6 stroke grestore} bind def
/D7 {gsave translate 45 rotate 0 0 S7 stroke grestore} bind def
/D8 {gsave translate 45 rotate 0 0 S8 stroke grestore} bind def
/D9 {gsave translate 45 rotate 0 0 S9 stroke grestore} bind def
/D10 {gsave translate 45 rotate 0 0 S10 stroke grestore} bind def
/D11 {gsave translate 45 rotate 0 0 S11 stroke grestore} bind def
/D12 {gsave translate 45 rotate 0 0 S12 stroke grestore} bind def
/D13 {gsave translate 45 rotate 0 0 S13 stroke grestore} bind def
/D14 {gsave translate 45 rotate 0 0 S14 stroke grestore} bind def
/D15 {gsave translate 45 rotate 0 0 S15 stroke grestore} bind def
/DiaE {stroke [] 0 setdash vpt add M
  hpt neg vpt neg V hpt vpt neg V
  hpt vpt V hpt neg vpt V closepath stroke} def
/BoxE {stroke [] 0 setdash exch hpt sub exch vpt add M
  0 vpt2 neg V hpt2 0 V 0 vpt2 V
  hpt2 neg 0 V closepath stroke} def
/TriUE {stroke [] 0 setdash vpt 1.12 mul add M
  hpt neg vpt -1.62 mul V
  hpt 2 mul 0 V
  hpt neg vpt 1.62 mul V closepath stroke} def
/TriDE {stroke [] 0 setdash vpt 1.12 mul sub M
  hpt neg vpt 1.62 mul V
  hpt 2 mul 0 V
  hpt neg vpt -1.62 mul V closepath stroke} def
/PentE {stroke [] 0 setdash gsave
  translate 0 hpt M 4 {72 rotate 0 hpt L} repeat
  closepath stroke grestore} def
/CircE {stroke [] 0 setdash 
  hpt 0 360 arc stroke} def
/Opaque {gsave closepath 1 setgray fill grestore 0 setgray closepath} def
/DiaW {stroke [] 0 setdash vpt add M
  hpt neg vpt neg V hpt vpt neg V
  hpt vpt V hpt neg vpt V Opaque stroke} def
/BoxW {stroke [] 0 setdash exch hpt sub exch vpt add M
  0 vpt2 neg V hpt2 0 V 0 vpt2 V
  hpt2 neg 0 V Opaque stroke} def
/TriUW {stroke [] 0 setdash vpt 1.12 mul add M
  hpt neg vpt -1.62 mul V
  hpt 2 mul 0 V
  hpt neg vpt 1.62 mul V Opaque stroke} def
/TriDW {stroke [] 0 setdash vpt 1.12 mul sub M
  hpt neg vpt 1.62 mul V
  hpt 2 mul 0 V
  hpt neg vpt -1.62 mul V Opaque stroke} def
/PentW {stroke [] 0 setdash gsave
  translate 0 hpt M 4 {72 rotate 0 hpt L} repeat
  Opaque stroke grestore} def
/CircW {stroke [] 0 setdash 
  hpt 0 360 arc Opaque stroke} def
/BoxFill {gsave Rec 1 setgray fill grestore} def
/Density {
  /Fillden exch def
  currentrgbcolor
  /ColB exch def /ColG exch def /ColR exch def
  /ColR ColR Fillden mul Fillden sub 1 add def
  /ColG ColG Fillden mul Fillden sub 1 add def
  /ColB ColB Fillden mul Fillden sub 1 add def
  ColR ColG ColB setrgbcolor} def
/BoxColFill {gsave Rec PolyFill} def
/PolyFill {gsave Density fill grestore grestore} def
/h {rlineto rlineto rlineto gsave fill grestore} bind def
%
% PostScript Level 1 Pattern Fill routine for rectangles
% Usage: x y w h s a XX PatternFill
%	x,y = lower left corner of box to be filled
%	w,h = width and height of box
%	  a = angle in degrees between lines and x-axis
%	 XX = 0/1 for no/yes cross-hatch
%
/PatternFill {gsave /PFa [ 9 2 roll ] def
  PFa 0 get PFa 2 get 2 div add PFa 1 get PFa 3 get 2 div add translate
  PFa 2 get -2 div PFa 3 get -2 div PFa 2 get PFa 3 get Rec
  gsave 1 setgray fill grestore clip
  currentlinewidth 0.5 mul setlinewidth
  /PFs PFa 2 get dup mul PFa 3 get dup mul add sqrt def
  0 0 M PFa 5 get rotate PFs -2 div dup translate
  0 1 PFs PFa 4 get div 1 add floor cvi
	{PFa 4 get mul 0 M 0 PFs V} for
  0 PFa 6 get ne {
	0 1 PFs PFa 4 get div 1 add floor cvi
	{PFa 4 get mul 0 2 1 roll M PFs 0 V} for
 } if
  stroke grestore} def
/languagelevel where
 {pop languagelevel} {1} ifelse
 2 lt
	{/InterpretLevel1 true def}
	{/InterpretLevel1 Level1 def}
 ifelse
%
% PostScript level 2 pattern fill definitions
%
/Level2PatternFill {
/Tile8x8 {/PaintType 2 /PatternType 1 /TilingType 1 /BBox [0 0 8 8] /XStep 8 /YStep 8}
	bind def
/KeepColor {currentrgbcolor [/Pattern /DeviceRGB] setcolorspace} bind def
<< Tile8x8
 /PaintProc {0.5 setlinewidth pop 0 0 M 8 8 L 0 8 M 8 0 L stroke} 
>> matrix makepattern
/Pat1 exch def
<< Tile8x8
 /PaintProc {0.5 setlinewidth pop 0 0 M 8 8 L 0 8 M 8 0 L stroke
	0 4 M 4 8 L 8 4 L 4 0 L 0 4 L stroke}
>> matrix makepattern
/Pat2 exch def
<< Tile8x8
 /PaintProc {0.5 setlinewidth pop 0 0 M 0 8 L
	8 8 L 8 0 L 0 0 L fill}
>> matrix makepattern
/Pat3 exch def
<< Tile8x8
 /PaintProc {0.5 setlinewidth pop -4 8 M 8 -4 L
	0 12 M 12 0 L stroke}
>> matrix makepattern
/Pat4 exch def
<< Tile8x8
 /PaintProc {0.5 setlinewidth pop -4 0 M 8 12 L
	0 -4 M 12 8 L stroke}
>> matrix makepattern
/Pat5 exch def
<< Tile8x8
 /PaintProc {0.5 setlinewidth pop -2 8 M 4 -4 L
	0 12 M 8 -4 L 4 12 M 10 0 L stroke}
>> matrix makepattern
/Pat6 exch def
<< Tile8x8
 /PaintProc {0.5 setlinewidth pop -2 0 M 4 12 L
	0 -4 M 8 12 L 4 -4 M 10 8 L stroke}
>> matrix makepattern
/Pat7 exch def
<< Tile8x8
 /PaintProc {0.5 setlinewidth pop 8 -2 M -4 4 L
	12 0 M -4 8 L 12 4 M 0 10 L stroke}
>> matrix makepattern
/Pat8 exch def
<< Tile8x8
 /PaintProc {0.5 setlinewidth pop 0 -2 M 12 4 L
	-4 0 M 12 8 L -4 4 M 8 10 L stroke}
>> matrix makepattern
/Pat9 exch def
/Pattern1 {PatternBgnd KeepColor Pat1 setpattern} bind def
/Pattern2 {PatternBgnd KeepColor Pat2 setpattern} bind def
/Pattern3 {PatternBgnd KeepColor Pat3 setpattern} bind def
/Pattern4 {PatternBgnd KeepColor Landscape {Pat5} {Pat4} ifelse setpattern} bind def
/Pattern5 {PatternBgnd KeepColor Landscape {Pat4} {Pat5} ifelse setpattern} bind def
/Pattern6 {PatternBgnd KeepColor Landscape {Pat9} {Pat6} ifelse setpattern} bind def
/Pattern7 {PatternBgnd KeepColor Landscape {Pat8} {Pat7} ifelse setpattern} bind def
} def
%
%
%End of PostScript Level 2 code
%
/PatternBgnd {
  TransparentPatterns {} {gsave 1 setgray fill grestore} ifelse
} def
%
% Substitute for Level 2 pattern fill codes with
% grayscale if Level 2 support is not selected.
%
/Level1PatternFill {
/Pattern1 {0.250 Density} bind def
/Pattern2 {0.500 Density} bind def
/Pattern3 {0.750 Density} bind def
/Pattern4 {0.125 Density} bind def
/Pattern5 {0.375 Density} bind def
/Pattern6 {0.625 Density} bind def
/Pattern7 {0.875 Density} bind def
} def
%
% Now test for support of Level 2 code
%
Level1 {Level1PatternFill} {Level2PatternFill} ifelse
/Symbol-Oblique /Symbol findfont [1 0 .167 1 0 0] makefont
dup length dict begin {1 index /FID eq {pop pop} {def} ifelse} forall
currentdict end definefont pop
end
gnudict begin
gsave
0 0 translate
0.050 0.050 scale
0 setgray
newpath
1.000 UL
LTb
900 600 M
63 0 V
4457 0 R
-63 0 V
900 1138 M
63 0 V
4457 0 R
-63 0 V
900 1675 M
63 0 V
4457 0 R
-63 0 V
900 2213 M
63 0 V
4457 0 R
-63 0 V
900 2750 M
63 0 V
4457 0 R
-63 0 V
900 3288 M
63 0 V
4457 0 R
-63 0 V
900 600 M
0 63 V
0 2625 R
0 -63 V
2030 600 M
0 63 V
0 2625 R
0 -63 V
3160 600 M
0 63 V
0 2625 R
0 -63 V
4290 600 M
0 63 V
0 2625 R
0 -63 V
5420 600 M
0 63 V
0 2625 R
0 -63 V
900 3288 M
900 600 L
4520 0 V
0 2688 V
-4520 0 V
stroke
LCb setrgbcolor
LTb
LCb setrgbcolor
LTb
1.000 UP
1.000 UL
LTb
3.000 UL
LT0
946 601 M
45 3 V
46 4 V
46 6 V
45 7 V
46 9 V
46 9 V
45 11 V
46 13 V
46 13 V
45 14 V
46 15 V
46 16 V
45 17 V
46 17 V
46 19 V
45 19 V
46 19 V
45 21 V
46 21 V
46 21 V
45 22 V
46 22 V
46 23 V
45 24 V
46 23 V
46 24 V
45 25 V
46 25 V
46 25 V
45 26 V
46 26 V
46 26 V
45 26 V
46 27 V
46 27 V
45 27 V
46 28 V
46 28 V
45 28 V
46 28 V
46 28 V
45 29 V
46 28 V
46 29 V
45 29 V
46 29 V
46 30 V
45 29 V
46 30 V
45 29 V
46 30 V
46 30 V
45 30 V
46 30 V
46 30 V
45 31 V
46 30 V
46 31 V
45 30 V
46 31 V
46 31 V
45 30 V
46 31 V
46 31 V
45 31 V
46 31 V
46 31 V
45 32 V
46 31 V
46 31 V
45 31 V
46 32 V
46 31 V
45 32 V
46 31 V
46 32 V
45 31 V
46 32 V
46 31 V
45 32 V
46 32 V
45 31 V
46 32 V
46 32 V
45 32 V
46 31 V
46 32 V
45 32 V
46 32 V
46 32 V
45 32 V
46 32 V
46 31 V
45 32 V
46 32 V
46 32 V
45 32 V
46 32 V
stroke
1.000 UL
LTb
900 3288 M
900 600 L
4520 0 V
0 2688 V
-4520 0 V
1.000 UP
stroke
grestore
end
showpage
  }}%
  \put(3160,100){\makebox(0,0){\strut{}$E/m$}}%
  \put(200,1944){%
%
  \makebox(0,0){\strut{}$F_1$}%
%
  }%
  \put(5420,400){\makebox(0,0){\strut{}3.0}}%
  \put(4290,400){\makebox(0,0){\strut{}2.5}}%
  \put(3160,400){\makebox(0,0){\strut{}2.0}}%
  \put(2030,400){\makebox(0,0){\strut{}1.5}}%
  \put(900,400){\makebox(0,0){\strut{}1.0}}%
  \put(780,3288){\makebox(0,0)[r]{\strut{}5.0}}%
  \put(780,2750){\makebox(0,0)[r]{\strut{}4.0}}%
  \put(780,2213){\makebox(0,0)[r]{\strut{}3.0}}%
  \put(780,1675){\makebox(0,0)[r]{\strut{}2.0}}%
  \put(780,1138){\makebox(0,0)[r]{\strut{}1.0}}%
  \put(780,600){\makebox(0,0)[r]{\strut{}0.0}}%
\end{picture}%
\endgroup
\end{center}
\caption[]{Variation of the rate with the energy of the incoming
neutrino for a charge symmetric medium.
\label{f:F1V}}
\end{figure}

As we have already mentioned, we have neglected the
electron mass in the calculation of the amplitude.
Therefore, the results that we have obtained are applicable
in situations in which the electron gas is extremely relativistic.
In particular, the following the hierarchical relationship
\begin{eqnarray}
\label{hierarchy}
T,\mu \gg m_e \gg m > m^\prime
\end{eqnarray}
must hold among the relevant parameters involved, where ``$T,\mu$'' on
the left-hand side means ``either $T$ or $\mu$, or both''. 
However, the final formulas still depend on some parameters including
the initial neutrino velocity and the state of the electron gas.
To give an idea of how the rate depends on them,
we consider two extreme cases regarding the values of $T$ and $\mu$ in
\Eq{n&rho}. For these purposes we write the formula for the decay rate
in the form
\begin{eqnarray}
\Gamma = \frac{\omega_0}{2\pi}
\left( {g^2 \kappa m} |U_{e1}U_{e2}^*| 
\right)^2 \left(\frac{C_1}{M^8_W}\right)F(V) \,,
\end{eqnarray}
with
\begin{eqnarray}
\label{F}
F(V) \equiv {1 \over V C_1} \sum_{n=0}^4 \alpha_n R_n\,.
\end{eqnarray}
%

%%%%%
\subsubsection{Charge-symmetric medium}
%%%%%
We consider first a medium with zero chemical potential.
As \Eq{n&rho} shows, in this case 
\begin{eqnarray}
A_{11} = 0 \,, \qquad A_{20} = - {7\pi^2 \over 720} T^4 \,.
%\label{}
\end{eqnarray}
Looking at \Eq{C1C2}, we find that $C_2$ vanishes in this case,
whereas $C_1$ is given by
\begin{eqnarray}
C_1 =  {49\pi^4 \over 57600} T^8 \,.
%\label{}
\end{eqnarray}
The rate of the gravitational decay is then given by
\begin{eqnarray}
\label{Ratemu=0case}
\Gamma = {49\pi^3 \over 115200}\omega_0
\left( {g^2 \kappa m} |U_{e1}U_{e2}^*| 
\right)^2 \left( {T \over M_W} \right)^8 F_1 \,,
\end{eqnarray}
where $F_1$ is given by \Eq{F} but setting $C_2 = 0$.
Furthermore, in this case the angular distribution of the gravitons
has the specific form
\begin{eqnarray}
\frac{1}{\Gamma}\frac{d\Gamma}{d\cos\theta} \propto
\frac{\sin^4\theta}{(1 - V\cos\theta)^2}\,,
\end{eqnarray}
which depends on the velocity of the initial neutrino, but
is independent of any other parameters.

Fig.~\ref{f:F1V} shows the plot of $F_1$ as a function of $E/m$.
For low energy neutrinos, $F_1 \rightarrow 0$ as $E \rightarrow m$,
as already commented. For high energy ($E \gg m$) neutrinos,
\begin{eqnarray}
\alpha_1 & \simeq & -4\left(\frac{m}{E}\right)C_1\,,\nonumber\\
\alpha_2 & \simeq & 4C_1 \,,\nonumber\\
\alpha_3 & \simeq & -4\left(\frac{m}{E}\right)C_1\,,\nonumber\\
\alpha_4 & \simeq & \left(\frac{m}{E}\right)^2 C_1\,,
\end{eqnarray}
while
\begin{eqnarray}
\label{RER}
R_1 & \simeq & 2\log\left(\frac{E}{m}\right)\,,\nonumber\\
R_n & \simeq & \frac{2^{n - 1}}{n - 1}\left(\frac{E}{m}\right)^{n - 1}\,,
\qquad(n \not= 1)
\end{eqnarray}
and therefore $F_1$ grows as
\begin{eqnarray}
\label{FER}
F_1 \simeq F^{(ER)} \equiv \frac{8E}{3m} \,.
\end{eqnarray}
The result for the rate remains valid as long as the relationship
$E \ll M_W$ is maintained.

%%%%%
\subsubsection{Completely degenerate medium}
%%%%%
We now consider the limit $T=0$. For this case \Eq{n&rho} gives
\begin{eqnarray}
\label{Afordegeneratecase}
A_{11} = {\mu^3 \over 12\pi^2} \,, \qquad 
A_{20} = - {\mu^4 \over 48\pi^2}  \,,
\end{eqnarray}
which in turn yield
\begin{eqnarray}
\label{C1C2deg}
C_1 &=& {\mu^8 \over 2^8 \times 9\pi^4} (3 + 4E/\mu)^2\,, \nonumber\\
\frac{C_2}{C_1} & = & \frac{-4\omega_0}{4E + 3\mu} \,.
\end{eqnarray}
However, notice that the relations in \Eq{hierarchy} imply
in this case that
\begin{eqnarray}
\left|\frac{C_2}{C_1}\right| \ll 1\,,
\end{eqnarray}
for all the values of the parameters consistent with the assumptions
that we have made. Therefore, the decay rate in this case is given by
\begin{eqnarray}
\Gamma = \frac{\omega_0}{2^9 \times 9\pi^5}
\left( {g^2 \kappa m} |U_{e1}U_{e2}^*|\right)^2
\left({\mu \over M_W} \right)^8 (3 + 4E/\mu)^2 F_1\,,
\end{eqnarray}
where $F_1$ is the same function that appears in \Eq{Ratemu=0case}.
Therefore in this case the rate grows proportional to $E$ for $E \ll
\mu$, or $E^3$ for $E \gg \mu$.

%
% $Id: amvdk_sec4.tex,v 1.1 2009/06/02 15:07:26 nieves Exp nieves $
%
% sec 4
%
\section{Concluding remarks}
\label{s:disc}

These two example calculations explicitly confirm the suggestion
that angular momentum violating processes can occur in a medium
which is completely isotropic, even though the state of the medium
does not change at all in the process. We can understand
intuitively how this happens as follows. To elaborate this argument
we use a radiative decay process, like the one described in
Sec.\,\ref{s:raddk}, as an example.

Consider the decay of a particle $\psi$ to a particle $\psi^\prime$
and a photon in the rest frame of the decaying particle.  Denoting the
total angular momentum of the initial and final states by $\vec J$ and
$\vec J\,'$ respectively, in the vacuum we have
\begin{eqnarray}
\vec J &=& \vec L_\psi + \vec S_\psi \,, \nonumber\\* 
\vec J\,' &=& \vec L_{\psi'} + \vec L_\gamma + \vec S_{\psi'}
+ \vec S_\gamma \,,
%\label{}
\end{eqnarray}
where $L_x$ and $S_x$ stand for orbital angular momentum and spin of
each particle. Angular momentum conservation implies in
particular that
\begin{eqnarray}
\langle\vec J\cdot\hat q\rangle = \langle\vec J^{\,\prime}\cdot\hat
q\rangle\,, 
\end{eqnarray}
where the brackets in the left and right-hand sides denote the
expectation value with respect to the initial and final states,
respectively.  However, since $\vec p = 0$, the orbital angular
momentum of the decaying particle is zero and, since and $\vec q = -
\vec p\,'$, neither $\vec L_{\psi^\prime}$ nor $\vec L_{\gamma}$ have
a component along $\hat q$. Therefore, conservation of angular
momentum implies
\begin{eqnarray}
\label{helicitycond}
\langle\vec S_\psi \cdot \hat q\rangle =
\langle\vec S_{\psi'}\cdot\hat q + \vec S_{\gamma}\cdot\hat q\rangle \,,
%\label{}
\end{eqnarray}
so that the helicity along the direction of motion of the final
particles is conserved in this frame.  Since the helicity can only be
zero for the scalars and $\pm1$ for the photon, the process involving
scalars is forbidden.

The diagrams of Fig.~\ref{f:diags} in which one of the fermion lines
denotes the thermal on-shell part of the fermion propagator,
physically corresponds to the process
\begin{eqnarray}
\label{realprocess}
f(l) + \psi(p) \rightarrow f(l) + \psi'(p') + \gamma(q) \,,
%\label{}
\end{eqnarray}
where $f$ is the fermion in the background medium. Then in place
of \Eq{helicitycond}, the relevant condition is now
\begin{eqnarray}
\langle\vec L_f\cdot\hat q + \vec S_f\cdot\hat q +
\vec S_\psi\cdot\hat q\rangle = 
\langle\vec L_{f}\cdot\hat q + \vec S_{f}\cdot\hat q +
\vec S_{\psi'}\cdot\hat q + \vec S_\gamma\cdot\hat q\rangle\,,
%\label{}
\end{eqnarray}
where the various terms on either side of this equation should of
course be added according to the angular momentum addition rules. Even
assuming that we may drop the spin contributions of the background
fermions upon averaging over an unpolarized background, their orbital
angular momentum does not vanish because in the rest frame of the
decaying particle the fermions appear to be moving in a preferred
direction.  Therefore, helicity is not a good quantum number anymore
and the helicity argument does not apply.  The argument can be
extended with minor modifications to the gravitational decay as well.

It should be emphasized that, while the model of Sec.\,\ref{s:raddk}
is a \emph{toy} model, in Sec.\,\ref{s:gravdk} we have worked with
nothing else than the standard model augmented with the linearized
gravitational couplings.  Thus, the effects calculated in
Sec.\,\ref{s:gravdk} constitute real predictions of the standard model,
subject to the approximations we have made. The fact that these processes
are forbidden in the vacuum by angular momentum conservation arguments,
implies that, in the medium, their angular distribution and differential
decay rates have a distinctive form. This could lead to observable
consequences in specific physical contexts despite the fact that
there may exist other competing processes with comparable total rates.

\end{document}